Title: Bubble Gating Currents in Ionic Channels

Authors: Bob Eisenberg

Bubbles in ion channel proteins have been proposed to be the bistable gates that control current flow. Gating currents associated with channel gating would then be an electrical signature of bubble breaking and formation, arising from the change in dielectric coefficient as the bubble breaks or forms. A bubble would have a dielectric coefficient of 1. A filled bubble would have a dielectric coefficient (say) between 30 and 80. Transporters, pumps, and channels would be expected to have gating currents.

# Bubble Gating Currents in Ion Channel Proteins
Version 2

Feb 8, 2008
Bob Eisenberg
beisenbe@rush.edu

Department of Molecular Biophysics and Physiology
Rush University Medical Center
1653 West Congress Parkway
Chicago IL 60612
USA

A recent paper (1) suggests that bubbles in ionic channels are the bistable hydrophobic gates that control the on-off switching of single channel currents. Here, I point out that gating currents may be an electrical signature of bubble breaking and formation.

When a bubble breaks inside the pore of a channel, the dielectric coefficient of the region containing the bubble will change substantially, from more or less one, to a much larger number, between 30 and 80 in all likelihood. (The dielectric coefficient inside a channel is unknown but is thought to be reduced below that of bulk water — ~80 — because of restrictions on the mobility of water and other solvating moieties, at least in the narrow regions of the channel). The change in dielectric coefficient inside the pore would produce a change in the capacitance of the membrane in which the pore and channel are embedded. The change in capacitance clearly would have many, if not all of the properties of gating current (2-6). Gating currents are known to be associated with the opening and closing of voltage dependent ionic channels.

Thus, I propose that bubble formation and breaking produce a change in capacitive (i.e., displacement) charge that can be observed as a bubble-gating current, a 'bubble current' for short. The part of this bubble-current that depends linearly on the applied potential would not be visible in the traditional measurements of channel gating current (in voltage dependent channels). The traditional experimental protocol subtracts currents measured for two different voltage steps and thus does not show gating (or ionic) currents that are proportional to voltage (at every time with the same proportionality constant). Such linear currents could be the result of bubble breaking, for example, and depend on time, but they would not be seen in traditionally processed records of gating current. The tiny early component of gating current seen in voltage activated channels (7) might well represent the movements of the charge on voltage sensor that moves and thus initiates and controls the breaking of bubbles. Ref (1) proposed that sensors, including voltage sensors, work by modulating bubble breaking and formation.

If the gating currents observed in nerve axons (for example) are the electrical signature of bubbles, bubble-gating currents should also be found in all channels and transporters that have bistable single channel currents (or fluxes) that suddenly switch from off to on, and then on to off. Bubble-gating currents should be seen upon sudden activation of agonist activated channels, of light activated channels, and of transporters, if this idea is correct. Activation gating currents have already been seen in many systems, for example, (8-11)

The hypothesis that bubble-gating currents are an electrical signature of bubble gating suggests that gas anesthetics should have significant effects on gating currents in all these channels and transporters. although the effects would not be expected to all be the same.